\documentclass[pre,twocolumn,amsmath,amssymb,showpacs]{revtex4}
\usepackage{color}
\usepackage{graphicx}

\newcommand{\diff}{{\mathrm{d}}}
\newcommand{\eh}{{\mathrm{e}^h}}
\begin{document}

\title{Stability of velocity-Verlet- and Liouville-operator-derived algorithms to integrate non-Hamiltonian systems}

\author{Hiroshi Watanabe}
\email{hwatanabe@issp.u-tokyo.ac.jp}
\thanks{Corresponding author}
\affiliation{%
The Institute for Solid State Physics, The University of Tokyo,
Kashiwanoha 5--1--5, Kashiwa, Chiba 277--8581, Japan
}

\begin{abstract}
We investigate the difference between the velocity Verlet and the Liouville-operator-derived (LOD) algorithms by studying two non-Hamiltonian systems, one dissipative and the other conservative, for which the Jacobian of the transformation can be determined exactly.
For the two systems, we demonstrate that (1) the velocity Verlet scheme fails to integrate the former system while the first- and second-order LOD schemes succeed, (2) some first-order LOD fails to integrate the latter system while the velocity Verlet and the other first- and second-order schemes succeed. We have shown that the LOD schemes are stable for the former system by determining the explicit forms of the shadow Hamiltonians which are exactly conserved by the schemes. We have shown that Jacobian of the velocity Verlet scheme for the former system and that of the first-order LOD scheme for the latter system are always smaller than the exact values, and therefore, the schemes are unstable. The decomposition-order dependence of LOD schemes is also considered.
\end{abstract}

\maketitle

\section{Introduction}

Since the pioneering work by Alder and Wainwright~\cite{Alder1957},
the molecular dynamics (MD) simulation has been a popular and one of the most important simulation tools
in various fields~\cite{Allen_1989}.
While the first MD simulation involved only 32 atoms, recent MD simulations have involved billions of atoms owing to the increase in computational power~\cite{fx10full}.
Since the computational power is still increasing, it is natural to want to perform MD simulations with a larger number of atoms and time steps.
The time evolution in MD simulations is performed numerically by integrating Newton's equations of motion.
In 1967, Verlet proposed an efficient method of integrating the equations of motion~\cite{Verlet1967}, which method was modified by Swope \textit{et al.}~\cite{Swope1982}.
The original and modified methods are referred to the position Verlet and velocity Verlet algorithms, respectively.
The velocity Verlet algorithm has been widely used since it is simple and achieves stable long-time integration.
The position and velocity Verlet algorithms were later found to be symplectic integrators.
It is known that a symplectic integrator conserves the Hamiltonian and therefore achieves stable integration over a long time. This is because there exists a value that is exactly conserved by the
approximated propagator~\cite{Tuckerman2006}. The conserved value is sometimes called a shadow Hamiltonian
since it is closely related to the original Hamiltonian.
The error between the shadow Hamiltonian and the original Hamiltonian is bounded by the time step $h$, and they become identical when $h \rightarrow 0$.
An explicit symplectic integrator is constructed by the decomposition of the Liouville operator associated with the equations of motion~\cite{Yoshida1990, Suzuki1991}. The velocity Verlet algorithm is equivalent to a symplectic integrator with second-order symmetric decomposition. While higher-order symplectic integrators can be constructed~\cite{Yoshida1990}, the second-order method is usually used.

While the Hamiltonian dynamics achieve the microcanonical ($NVE$) ensemble provided the system is ergodic,
the canonical ($NVT$) ensemble is often required since the temperature dependence of the observables is usually of interest rather than their total-energy dependence.
To achieve the canonical ensemble, a thermostat, such as the Nos\'e--Hoover method~\cite{Hoover1985} and its family, is widely used~\cite{Martyna1992}.
Tuckerman \textit{et al.} constructed a new integration scheme for non-Hamiltonian dynamics
by the decomposition of the Liouville operator similarly to a symplectic integrator for Hamiltonian dynamics~\cite{Tuckerman1990, Tuckerman1992}.
Martyna \textit{et al.} developed explicit integrators which are reversible and yields the isothermal and/or isobaric ensembles~\cite{Martyna1996}.
The construction of a numerical integrator by the decomposition of the Liouville operator
was later applied to a system with a stochastic thermostat~\cite{Leimkuhler2013}.

The velocity Verlet algorithm is sometimes identified with the second-order symplectic integrator.
Actually, they are identical when applied to Hamiltonian dynamics. However, the original velocity Verlet formula
can be applied to non-Hamiltonian dynamics, which leads to a different scheme from that constructed by the decomposition of the Liouville operator. Thus, a natural question arises; which is better?
The suitability of the time integrator for non-Hamiltonian dynamics has mainly been investigated from the viewpoint of the error, \textit{i.e.}, the time-step or system size dependence of the deviation from the expected value~\cite{Leimkuhler2013, Itoh2013}.
However, the difference between the two is not limited to the integration accuracy. One exhibits unstable integration for some systems while the others allow successful integration and vice versa.
For example, the velocity-Verlet-based algorithm exhibits secular growth in the conserved quantity for long-time simulation while the appropriate algorithms exhibit stable integrations~\cite{Tuckerman1999-2}.

It has been shown that the LOD algorithms conserve the volume of the phase space,
in other words, the Jacobian of the transformation~\cite{Tuckerman2006}. However, it is usually difficult to
determine the exact Jacobian for general dynamics.
In the present article, we consider two non-Hamiltonian systems for which the Jacobian can be exactly determined.
Since we know the exact form of the Jacobian, we can investigate the properties of the integrators more precisely beyond the error analysis of the observables, which is the purpose of this study.

The rest of the article is organized as follows.
In the next section, we give brief derivations of 
the velocity-Verlet and the LOD algorithms.
We will consider two non-Hamiltonian systems, a dissipative system in and a conserved system. In Sec.~\ref{sec:jacobian}, We give discussions of shadow Hamiltonians and Jacobians of the system to consider the stability of the integration schemes.
In Sec.~\ref{sec:decomposition}, we study the decomposition-order dependence on the stability and the accuracy. Finally, Sec.~\ref{sec:summary} is devoted to a summary and discussion.

\section{Integration Algorithm}

\subsection{Velocity Verlet algorithm}

We first give a brief derivation of the velocity Verlet algorithm.
Consider the following equations of motion:
\begin{equation}
\left\{
\begin{array}{ll}
\dot{v} &= f, \\
\dot{r} &= v,
\end{array}
\right.\label{eq:motion}
\end{equation}
where $r, v$, and $f$ denote position, velocity, and force, respectively.
For simplicity, the mass is set to unity.
The integration scheme is a map from $(r(t), v(t))$ to $(r(t+h),v(t+h))$ where $t$ is the current time and $h$ is the time step.
We consider a conservative force, \textit{i.e.}, the force is expressed by a 
potential function as $f(r) = V'(r)$, then the total energy $H = v^2/2 + V(r)$ is conserved throughout the time evolution.
Therefore, the integration scheme conserving the total energy is favorable
and the velocity Verlet algorithm satisfies this condition.

The velocity Verlet algorithm is constructed as follows~\cite{Swope1982}.
First, the position at the next step $r(t+h)$ is given by
\begin{equation}
r(t+h) = r(t) + h v(t) + \frac{h^2}{2} f(t), \label{eq:r}
\end{equation}
which is simply the Tayler series up to the second order.
Next, we define the velocity at the next step by the central difference formula as
\begin{equation}
v(t+h) = \frac{r(t+2 h) + r(t)}{2 h }. \label{eq:v}
\end{equation}
Applying Eq.~(\ref{eq:r}) to Eq.~(\ref{eq:v}) twice, we have 
\begin{equation}
v(t+h) = v(t) + \frac{f(t) + f(t+h)}{2 h}. \label{eq:v2}
\end{equation}
Combining Eqs.~(\ref{eq:r}) and (\ref{eq:v2}), we have the velocity Verlet algorithm.
Note that the force at the next step $f(t+h)$ is required to calculate
the velocity at the next step $v(t+h)$.
When the force depends only on the position, we can construct an
explicit scheme, \textit{i.e.}, first we calculate $r(t+h)$, then
we calculate $f(t+h)$ using $r(t+h)$, and finally we calculate $v(t+h)$ using $f(t+h)$.
The case of a velocity-dependent force will be considered later.

\subsection{Liouville-operator-derived algorithms}

Next, we introduce the LOD integration algorithms.
As described above, the velocity Verlet algorithm is identical to the second-order LOD scheme
when it is applied to Hamiltonian systems.
The Liouville operator of the equations of motion in Eq.~(\ref{eq:motion}) is given by
\begin{equation}
i \mathcal{L} = f \frac{\partial}{\partial v} + v \frac{\partial}{\partial r}.
\end{equation}
Then the propagator $U(h)$, which proceeds the time by $h$, is expressed by
\begin{equation}
U(h) = \exp{(i h \mathcal{L})}.
\end{equation}
Since the exact form of the propagator cannot be obtained,
the approximated propagator is used for the integration.
We decompose the Liouville operator as $i\mathcal{L} = i\mathcal{L}_A +i\mathcal{L}_B$, where
\begin{align}
i\mathcal{L}_A &=  v \frac{\partial}{\partial r} \\
i\mathcal{L}_B &=  f \frac{\partial}{\partial v}.
\end{align}
Here and throughout the rest of work, we denote a Liouville operator acting on coordinates as $ih \mathcal{L}_A$ and that acting on momenta as $ih \mathcal{L}_B$, where $ih \mathcal{L}_A$ is so-called ``drift" and $ih \mathcal{L}_B$ is ``kick" terms in the leapfrog algorithm.
Then the propagator can be approximated by the decomposition as
\begin{align}
\tilde{U}_1(h) &= \mathrm{e}^{i h \mathcal{L}_B}\mathrm{e}^{i h \mathcal{L}_A} \\
\tilde{U}_2(h) &= \mathrm{e}^{i h \mathcal{L}_A/2}\mathrm{e}^{i h \mathcal{L}_B} \mathrm{e}^{i h \mathcal{L}_A/2},
\end{align}
where $\tilde{U}_1$ and $\tilde{U}_2$ are the first- and second-order approximations of the original propagator $U$, respectively.
The approximated propagators conserve the phase space volume, \textit{i.e}.,
the Jacobian of the propagators is unity. This means that the propagators involve symplectic maps, and therefore, an integration scheme constructed by the decomposition of the Liouville operator is called a symplectic integrator.
Note that, the first-order LOD is rarely used for the practical use.
The reason why we consider the first-order LOD is to investigate the relation between the time-reversibility
and the stability of the integration. The 2nd-order LOD with the symmetric decomposition becomes time-reversible scheme, while the 1st-order LOD is always time-irreversible. In the following, we consider two systems, one is time-irreversible and the other is time-reversible, respectively. The relation of time-reversibility between the system and the scheme will be considered later.

Lastly, we show that the velocity Verlet algorithm is the symplectic integrator.
Consider the second-order symplectic integrator $\tilde{U}_2$. 
The integration scheme associated with $\tilde{U}_2$ is given by
\begin{align}
v(t+h/2) &= v + \frac{h f(t)}{2}, \\
r(t+h) &\leftarrow r(t) + h v(t+h/2),\\
v(t+h) &\leftarrow v(t+h/2) + \frac{h f(t+h)}{2}.
\end{align}
Eliminating $v(t+h/2)$, we have the velocity Verlet algorithm.
A different order of the decomposition leads to the position Verlet algorithm~\cite{Itoh2013}.

\section{Jacobian and Invariant Measure} \label{sec:jacobian}

\subsection{Theoretical Background}

We introduce brief arguments between Jacobian of transformation and the invariant measure of the phase space~\cite{Tuckerman1999, Tuckerman2001}.
For the simplicity, we consider a system having one degree of freedom.
It is straightforward to apply the following arguments to a system with any degree of freedoms.
Suppose the position and the velocity of the system at time $t$ are given by $r(t)$ and $v(t)$.
The coordinates of the phase space is denoted by $\textbf{x} \equiv (r,v)$.
Then the equations of motion is a map from the coordinates $x$ to the velocity field
$\dot{\textbf{x}} = (\dot{r}, \dot{v})$ in the phase space and time evolution of the system
is flow along the velocity field. After the time interval $h$, the state of the system is moved from $\textbf{x}(t) = (r,v)$ to $\textbf{x}(t+h) =  (R, V)$,
where $R \equiv r(t+h)$ and $V \equiv v(t+h)$, respectively.
The time evolution with a finite time interval involves the variable transformation 
from $(r,v)$ to $(R,V)$.
The Jacobian of this transformation from $t$ to $t+h$ is given by
\begin{align}
J(t + h; t) & = \frac{\partial (R,V)}{\partial (r,v)}\\
&= \frac{\partial R}{\partial r} \frac{\partial V}{\partial v}
- \frac{\partial R}{\partial v} \frac{\partial V}{\partial r}.
\end{align}
Taking time derivative, we have
\begin{equation}
\frac{\diff J}{\diff t} = \kappa J, \label{eq:dJ}
\end{equation}
where
\begin{equation}
\kappa \equiv \nabla \dot{\textbf{x}} =  \frac{\partial \dot{r}}{\partial r}
+ \frac{\partial \dot{v}}{\partial v}.
\end{equation}
Since the quantity $\kappa$ is the divergence of the velocity field in the phase space,
and therefore, it denotes the compressibility of the flow involved by the equations of motion.
Integrating Eq.~(\ref{eq:dJ}), we have
\begin{align}
J(t+h, t) &= \exp \left( \int_t^{t+h} \kappa \diff t \right), \\
&= \exp\left[ w(t+h) - w(t)\right], \label{eq:Ja}
\end{align}
where $w(t)$ satisfies the following condition
\begin{equation}
\dot{w} \equiv \kappa.
\end{equation}
The Jacobian connects the volume of the phase space as
\begin{equation}
\diff R \diff V = J \diff r \diff v. \label{eq:Jb}
\end{equation}
From Eqs.~(\ref{eq:Ja}) and (\ref{eq:Jb}), we have
\begin{equation}
\exp\left[-w(t+h)\right] \diff R \diff V = \exp\left[-w(t)\right] \diff r \diff v.
\end{equation}
The above equation means that the volume factor $\mathrm{e}^{-w} \diff \textbf{x}$ is conserved
throughout the time evolution and it plays the role of the invariant measure.
When the dynamics is governed by Hamiltonian $H(r,v)$, then the compressibility $\kappa$
becomes zero since
\begin{align}
\kappa &= \frac{\partial \dot{r}}{\partial r} + \frac{\partial \dot{v}}{\partial v} \\
&= \frac{\partial^2 H}{\partial r \partial v} - \frac{\partial^2 H}{\partial r \partial v} \\
&= 0.
\end{align}
Therefore, the Hamiltonian dynamics involves incompressible flow in the phase space.
As a result, the Jacobian becomes unity. Then the volume of the phase space is conserved as
\begin{equation}
\diff R \diff V = \diff r \diff v,
\end{equation}
which is nothing but the Liouville theorem.
As described above, the Jacobian associated with the symplectic integrator is exactly unity
and this is one of the reasons why the symplectic integrator allows us the stable integration.
Similarly, if an integration scheme conserves the exact value of Jacobian of a non-Hamiltonian system,
then the invariant measure $\mathrm{e}^{w} \diff \textbf{x}$ is also conserved throughout the time evolution.
Therefore, it is expected that the scheme which conserves the exact value of Jacobian is preferable.
In general, it is difficult to obtain the exact closed-form of the Jacobian for general non-Hamilton systems.
But in some cases, the exact form of the Jacobian can be determined without the exact solution of the equations of motion. In the following sections, we consider two systems whose Jacobian can be determined in the closed-form.

\subsection{Dissipative System}\label{sec:dissipative_system}

We will show that the velocity Verlet algorithm applied to a non-Hamiltonian system
is different from the schemes constructed by the decomposition of the Liouville operator.
Consider the following system.
\begin{equation}
\left\{
\begin{array}{ll}
\dot{v} &= -r -v \\
\dot{r} &= v
\end{array}
\right.\label{eq:dissipative}
\end{equation}
This system represents a harmonic oscillator with friction.
Note that, this equations of motion is time-irreversible, since it is not invariant for 
the transformation $v \rightarrow -v$ and $\dot{r} \rightarrow - \dot{r}$.
This system has a time-dependent conserved value given as
\begin{equation}
H = \mathrm{e}^{t}\left(
 r^2 + rv + v^2
\right). \label{eq:conserved_value1}
\end{equation}
One can confirm that $\dot{H} = 0$ by using the equations of motion in Eq.~(\ref{eq:dissipative}).
While Eq.~(\ref{eq:conserved_value1}) is not a Hamiltonian in a strict sense, we refer to $H$ as the Hamiltonian of this system for convenience. Similarly, we refer to a value exactly conserved by an approximated propagator as a shadow Hamiltonian.

To integrate this system, we consider three schemes, the velocity Verlet and first- and second-order LOD algorithms. The velocity Verlet algorithm for this system is given by
\begin{align}
R &= r + h v - \frac{h^2}{2}(r+v) \\
V &= v - \frac{h}{2}(r+v+R+V), \label{eq:VV}
\end{align}
where $R \equiv r(t+h)$ and $V \equiv v(t+h)$.
While the velocity at the next step $V$ appears on both sides of Eq.~(\ref{eq:VV}),
we can solve it with respect to $V$ as
\begin{equation}
V = \frac{2 v - h (r+v+R)}{2+h},
\end{equation}
then we obtain the explicit form of the integration scheme.

Hereinafter, we denote a non-Hermitian operator as $ih \mathcal{L}_O$.
While this label ``O" originates from the Ornstein-Uhlenbeck process in the Langevin dynamics~\cite{Leimkuhler2013}, we adopt this notation in the deterministic dynamics for the convenience.
The propagators of the first- and second-order LOD algorithms can be written as
\begin{align}
\tilde{U}_1(h) &= \mathrm{e}^{i h \mathcal{L}_A} \mathrm{e}^{i h \mathcal{L}_B} \mathrm{e}^{i h \mathcal{L}_O}, \label{eq:dissipative_lod1}\\
\tilde{U}_2(h) &= \mathrm{e}^{i h \mathcal{L}_B/2}\mathrm{e}^{i h \mathcal{L}_A /2} \mathrm{e}^{i h \mathcal{L}_O}\mathrm{e}^{i h \mathcal{L}_A /2}\mathrm{e}^{i h \mathcal{L}_B/2} ,
\end{align}
where
$i\mathcal{L}_A = v \partial_r, 
i\mathcal{L}_B = -r \partial_v$, and 
$i\mathcal{L}_O = v \partial_v$, respectively.
The second-order LOD algorithm exhibits the so-called BAOAB form~\cite{Gronbech-Jensen2013,Leimkuhler2013, Zhang2017}.
The operator $\exp(i h\mathcal{L}_O)$ is called the scaling operator~\cite{Tuckerman2006}, which yields
\begin{equation}
\exp(i h\mathcal{L}_O) v = v \mathrm{e}^{-h}.
\end{equation}
The integration procedure of $\tilde{U}_1(h)$ is given by
\begin{align}
r &\leftarrow r + v h,\\
v &\leftarrow v - r h, \\
v &\leftarrow v \mathrm{e}^{-h},
\end{align}
and that of $\tilde{U}_2(h)$ is
\begin{align}
v &\leftarrow v - r h/2, \\
r &\leftarrow r + v h/2,\\
v &\leftarrow v \mathrm{e}^{-h},\\
r &\leftarrow r + v h/2, \\
v &\leftarrow v - r h/2.
\end{align}

\begin{figure}
\includegraphics[width=7cm]{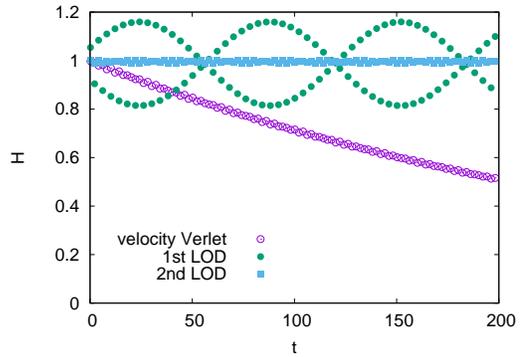}
\caption{(Color online) Time evolution of the Hamiltonian $H$ of Eq.~(\ref{eq:conserved_value1}).
The initial condition is $(v,r) = (0,1)$ and the time step is $0.2$.
The total number of steps is 1000 and the results every 10 steps are shown.
}\label{fig:dissipative}
\end{figure}

The time evolutions of the Hamiltonian $H$ integrated by the three schemes are shown in 
Fig.~\ref{fig:dissipative}.
While the results of the first- and the second-order LOD algorithms fluctuate around 1, 
that of the velocity Verlet algorithm decreases monotonically. These results
suggest that there exist shadow Hamiltonians that are exactly conserved in the LOD algorithms, whereas the velocity Verlet algorithm does not have such conserved values.
While it is difficult to find an exact closed-form expression for each shadow Hamiltonian,
we can find such expressions when the system is linear~(see Appendix A).
The shadow Hamiltonians are
\begin{align}
\tilde{H}_1 &= \mathrm{e}^t \left(
\mathrm{e}^h r^2 +
\frac{h^2 + \mathrm{e}^h - 1}{h} r v +  v^2\right),\\
\tilde{H}_2 &= \mathrm{e}^t \left[
r^2 +\frac{2(\eh -1)rv}{h(1+\eh)} + \left(1 - \frac{h^2}{4}\right)v^2
\right],
\end{align}
where $\tilde{H}_1$ and $\tilde{H}_2$ are exactly conserved by propagators $\tilde{U}_1$ and $\tilde{U}_2$, respectively.
It is straightforward to confirm that the errors between the shadow and original Hamiltonians are bounded as $\tilde{H}_1 - H = O(h)$ and $\tilde{H}_2 - H = O(h^2)$. The existence of the shadow Hamiltonians 
enables the LOD algorithms to achieve stable integration.

Next, we show why the velocity Verlet algorithm does not have a shadow Hamiltonian
such that the error from the original Hamiltonian is bounded.
Consider the time evolution of this system with the equations of motion in Eq.~(\ref{eq:dissipative}) from $t$ to $t + h$.
This time evolution involves a map of variables from $(r, v)$ to $(R, V$).
The compressibility of the phase flow associated with the equation of motion~(\ref{eq:dissipative}) is
\begin{equation}
\kappa = \frac{\partial \dot{r}}{\partial r} + \frac{\partial \dot{v}}{\partial v} = -1.
\end{equation}
If we consider the following value
\begin{equation}
\omega= \ln(v^2+ rv + r^2),
\end{equation}
then it satisfies the following identity
\begin{equation}
\dot{\omega} = -1 = \kappa.
\end{equation}
Therefore, the metric factor of this system is
\begin{equation}
\mathrm{e}^{-\omega} = \frac{1}{v^2+rv+r^2}.
\end{equation}
As shown in Eq.~(\ref{eq:Ja}), the exact Jacobian is given by
\begin{align}
J(t+h, t) &= \exp\left[\omega(t+h) - \omega(t) \right],\\
&= \frac{V^2+RV+R^2}{v^2+rv+r^2}. \label{eq:Jbefore}
\end{align}
From the conserved value in Eq.~(\ref{eq:conserved_value1}), we have the following identity
\begin{equation}
\mathrm{e}^{t+h} (V^2+RV+R^2) = \mathrm{e}^{t} (v^2+rv+r^2). \label{eq:conserved2}
\end{equation}
Inserting Eq.~(\ref{eq:conserved2}) into Eq.~(\ref{eq:Jbefore}), we have
\begin{equation}
J(t+h, t) = \mathrm{e}^{-h}.
\end{equation}
One can confirm that the Jacobians for the first- and the second-order LOD algorithms are both exactly $\mathrm{e}^{-h}$.
However, the Jacobian of the velocity Verlet algorithm is
\begin{equation}
\frac{\partial (V,R)}{\partial (v,r)} = \frac{2-h}{2+h},
\end{equation}
which is different from the exact value $\mathrm{e}^{-h}$.
Since $(2-h)/(2+h)$ is always smaller than $\exp(-h)$ for $h>0$, the volume of the phase space of the system integrated by the velocity Verlet algorithm is reduced smaller by $(2-h)\mathrm{e}^h/(2+h)$ every time step when it is compared with the volume of the phase space decrease of the system integrated by the LOD algorithms.
Consequently, the Hamiltonian $H$ in Eq.~(\ref{eq:conserved_value1}) also decreases at the same rate.

\begin{figure}
\includegraphics[width=7cm]{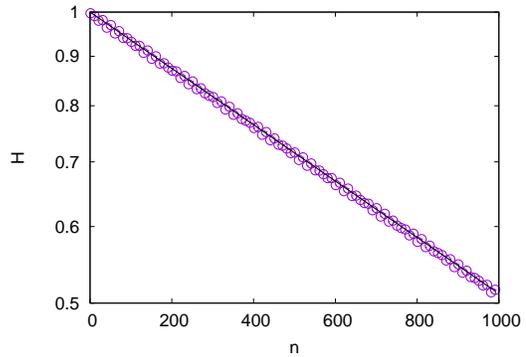}
\caption{Time evolution of the Hamiltonian $H$ of Eq.~(\ref{eq:dissipative}) integrated by the velocity Verlet algorithm as a function of time step $n$. The decimal logarithm is taken for the vertical axis.
The open circles are the values of $H$ integrated by the velocity Verlet algorithm
and the solid line denotes ${\left[(2-h)\mathrm{e}^h/(2+h)\right]}^n$ with
the time step $h=0.2$.
}\label{fig:decrease}
\end{figure}

Figure~\ref{fig:decrease} shows the time evolution of the Hamiltonian $H$ as a
function of the velocity Verlet step $n$. The rate of decrease is $(2-h)\mathrm{e}^h/(2+h)$ as predicted.
Since the Jacobian of the velocity Verlet algorithm is always smaller than the exact value,
the Hamiltonian decreases monotonically. Therefore, we can conclude that the velocity Verlet algorithm does not have a shadow Hamiltonian for which the error from the original Hamiltonian is bounded.

\subsection{Conserved System}\label{sec:conserved_system}

In the previous subsection, we considered a dissipative system.
Here, we investigate a non-Hamiltonian system with a time-independent conserved value.
Consider the system described by the following equations of motion:
\begin{equation}
\left\{
\begin{array}{ll}
\dot{v} &= - 2 r - 2 r v, \\
\dot{r} &= v.
\end{array}
\right.\label{eq:eom_conserved}
\end{equation}
This equations of motion are obtained by the adiabatic approximation of 
a harmonic oscillator with the Nos\'e--Hoover thermostat (see Appendix B).
Unlike the previous example, this equations of motion are time reversible.
One can obtain the following a conserved value of this system by the separation of variables:
\begin{equation}
H = r^2 + v - \ln(v+1). \label{eq:H_conserved}
\end{equation}
We also refer to $H$ as the Hamiltonian for convenience.
The existence of the conserved value causes a harmonic oscillator with the Nos\'e--Hoover thermostat
to lose its ergodicity~\cite{Legoll2006, Watanabe2007}.

While the equations of motion in Eq.~(\ref{eq:eom_conserved}) is not solvable, the Jacobian of this system can be obtained exactly.
Consider the time evolution from $t$ to $t+h$, which involves a map from
$(r,v)$ to $(R,V)$, where $R \equiv r(t+h)$ and $V \equiv v(t+h)$.
The Jacobian of this map is
\begin{align}
\frac{\partial (R,V)}{\partial (r,v)} &= \exp\left[\int_t^{t+h} \left(\frac{\partial \dot{r}}{\partial r} +  \frac{\partial \dot{v}}{\partial v}  \right) dt \right],  \nonumber \\
&= \exp\left[\int_t^{t+h} (-2r) dt \right],  \nonumber \\
&= \exp\left[\int_t^{t+h} (-2r)\frac{dt}{dv} dv \right],  \nonumber \\
&= \exp\left[\int_t^{t+h} \frac{dv}{v+1} \right], \nonumber \\
&= \frac{V+1}{v+1}. \label{eq:jacobian_nh_ad}
\end{align}

The velocity Verlet algorithm for this system can be obtained as
\begin{align}
R &= r + v h - r(v+1) h^2, \\
V &= \frac{v - h (r + R + rv)}{1+hR}.
\end{align}

The Liouville operator of this system is
$i \mathcal{L} = i \mathcal{L}_A+i \mathcal{L}_B + i \mathcal{L}_O$, where 
\begin{align}
i\mathcal{L}_A &= v \frac{\partial}{\partial r}, \\
i\mathcal{L}_B &= -2 r \frac{\partial}{\partial v},\\
i\mathcal{L}_O &= -2 r v \frac{\partial}{\partial v}. \\
\end{align}
While $i\mathcal{L}_A$ and $i\mathcal{L}_B$ are Hermitian, $i\mathcal{L}_O$
is non-Hermitian.
The operator $i\mathcal{L}_O$ yields the scaling operator
\begin{equation}
\exp(i h\mathcal{L}_O) v = v \mathrm{e}^{-2hr}.
\end{equation}
We consider another scaling operator $\exp\left[i h (\mathcal{L}_B + \mathcal{L}_O)\right]$, which yields
\begin{align}
\exp\left[i h (\mathcal{L}_B + \mathcal{L}_O)\right] v &= 
\exp\left[ - 2 r (v+1) \frac{\partial}{\partial v} \right] v \\
&= \mathrm{e}^{- 2 r h} (v + 1) - 1.
\end{align}
Using the above operators, we factorize the propagator in three ways as
\begin{align}
\tilde{U}_{1a} &= \mathrm{e}^{i h \mathcal{L}_A} \mathrm{e}^{i h \mathcal{L}_B} \mathrm{e}^{i h \mathcal{L}_O}, \label{eq:UABO}\\
\tilde{U}_{1b} &= \mathrm{e}^{i h \mathcal{L}_A} \mathrm{e}^{i h (\mathcal{L}_B + \mathcal{L}_O)}, \\
\tilde{U}_{2} &= \mathrm{e}^{i h \mathcal{L}_B/2} \mathrm{e}^{i h \mathcal{L}_A/2} \mathrm{e}^{i h \mathcal{L}_O} \mathrm{e}^{i h \mathcal{L}_A/2} \mathrm{e}^{i h \mathcal{L}_B/2},
\end{align}
where $\tilde{U}_{1a}$ and $\tilde{U}_{1a}$ are the first-order approximations and 
$\tilde{U}_{2}$ is the second-order approximation with the BAOAB form.
We omit the derivation of the integration procedures involving the above propagators since they are straightforward.

\begin{figure}
\includegraphics[width=7cm]{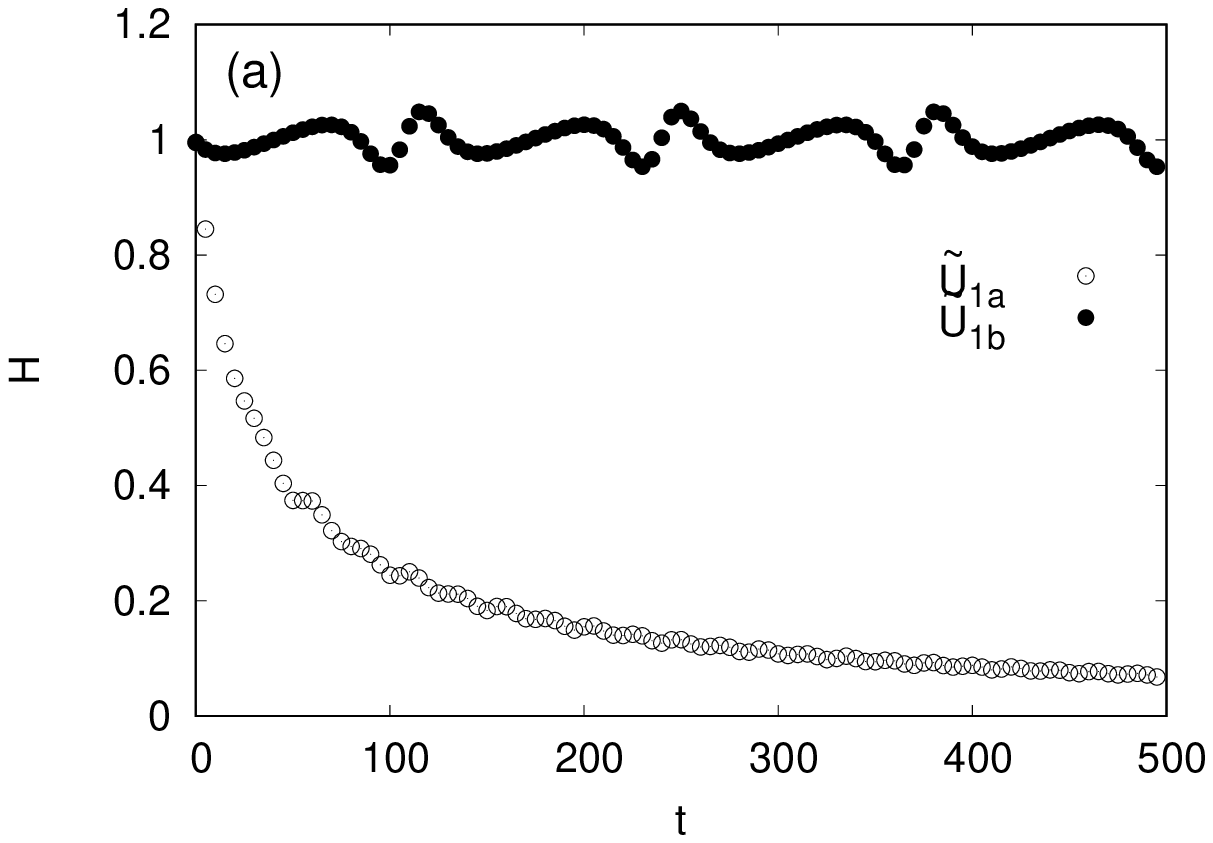}
\includegraphics[width=7cm]{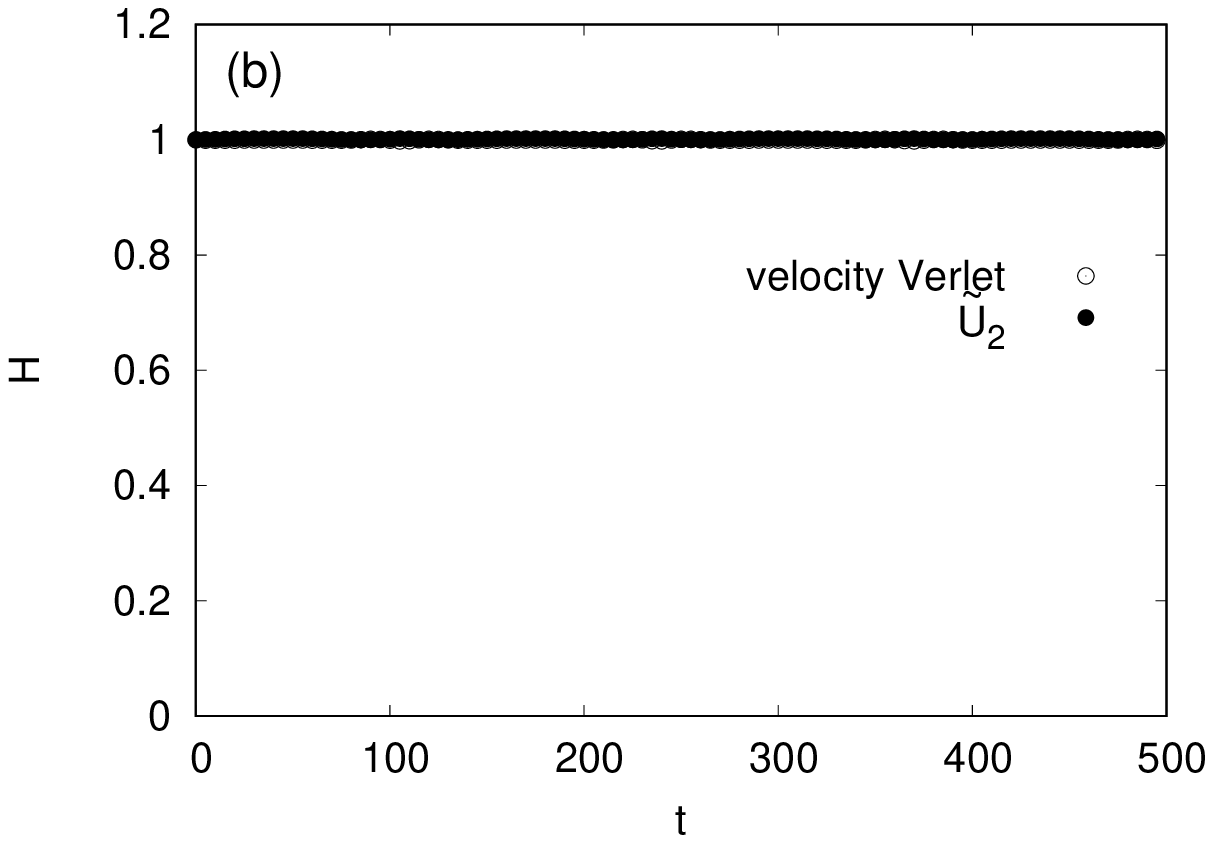}
\caption{Time evolution of the Hamiltonian $H$.
The results of $\tilde{U}_{1a}$ and $\tilde{U}_{1b}$ are shown in (a)
and those of the velocity Verlet schemes and $\tilde{U}_{2}$ are shown in (b).
The initial condition is $(v,r) = (0,1)$ and the time step is $0.2$.
The total number of steps is 10 000 and the results every 100 steps are shown.
}\label{fig:nh_ad}
\end{figure}

The time evolutions of the Hamiltonian for the propagators $\tilde{U}_{1a}$ and $U_{1b}$ are shown
in Fig.~\ref{fig:nh_ad}~(a). Although both propagators are first-order schemes, $\tilde{U}_{1b}$
conserves the Hamiltonian while $\tilde{U}_{1a}$ does not.
The absence of the shadow Hamiltonian for $\tilde{U}_{1a}$ can be proved by deriving 
the following inequality:
\begin{equation}
\frac{\partial (R,V)}{\partial (r,v)} < \frac{V+1}{v+1}. \label{eq:inequality}
\end{equation}
The explicit form of the map from $(r,v)$ to $(R,V)$ for $\tilde{U}_{1a}$ is
\begin{align}
V &= (v - 2 r h)\exp(- 2 r h), \\
R &= r + V h.
\end{align}
The Jacobian of this map is given by
\begin{equation}
\frac{\partial (R,V)}{\partial (r,v)} = \mathrm{e}^{- 2 r h},
\end{equation}
and the explicit form of $(V+1)/(v+1)$ is
\begin{equation}
\frac{V+1}{v+1} = \frac{1 + (v - 2rh)\mathrm{e}^{- 2 r h}}{1+v}.
\end{equation}
Consider the inequality
\begin{equation}
\mathrm{e}^{-x} < 1 - x \mathrm{e}^{-x},
\end{equation}
which holds for any value of $x$.
Replacing $x$ by $2 r h$, we have
\begin{equation}
\mathrm{e}^{-2 r h} < 1 - 2 r h \mathrm{e}^{-2 rh}.
\end{equation}
Adding $v \mathrm{e}^{- 2 r h}$ to both sides, we have
\begin{equation}
(1+v) \mathrm{e}^{-2 r h} < 1 + (v - 2 r h) \mathrm{e}^{-2 rh}. \label{eq:proof}
\end{equation}
Since the Hamiltonian has the form $H = r^2 + v - \ln(v+1)$, $v+1$ should always be positive.
Dividing both sides of Eq.~(\ref{eq:proof}) by $v+1$,
we have the inequality
\begin{equation}
\mathrm{e}^{-2 r h} < \frac{1 + (v - 2 r h) \mathrm{e}^{-2 rh}}{v+1},
\end{equation}
thus deriving Eq.~(\ref{eq:inequality}).
Since the Jacobian of $\tilde{U}_{1a}$ is always smaller than the exact value,
the Hamiltonian monotonically decreases as shown in Fig.~\ref{fig:nh_ad}.

\begin{figure}
\includegraphics[width=7cm]{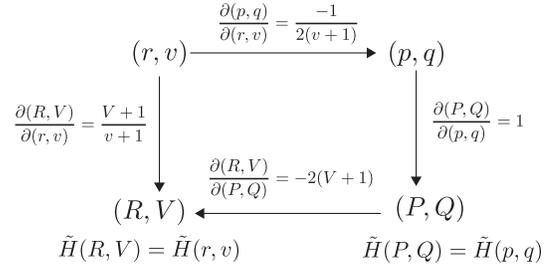}
\caption{Relation between $(r,v)$-space and $(p,q)$-space.
With the time evolution from $t$ to $t+h$, $(r,v)$ becomes $(R,V)$
and $(p,q)$ becomes $(P,Q)$.
}\label{fig:relation}
\end{figure}

While the propagator $\tilde{U}_{1b}$ is constructed by the first-order decomposition similarly to
$\tilde{U}_{1a}$, this propagator is designed so that it
yields the exact Jacobian.
The explicit form of the map from $(r,v)$ to $(R,V)$ for $\tilde{U}_{1b}$ is
\begin{equation}
\left\{
\begin{array}{ll}
V &= \mathrm{e}^{- 2 r h}(v+1) - 1,\\
R &= r + V h,
\end{array}
\right.\label{eq:vr}
\end{equation}
and one can confirm that the Jacobian of this map satisfies the
condition in Eq.~(\ref{eq:jacobian_nh_ad}) because
\begin{equation}
\frac{\partial (R,V)}{\partial (r,v)} =\mathrm{e}^{- 2 r h} = \frac{V+1}{v+1}.
\end{equation}
The fact that the Jacobian of this map is exact suggests that there
exists a shadow Hamiltonian for which the error from the original Hamiltonian is bounded.
However, it is difficult to find a closed-form expression for the shadow Hamiltonian since the system is
nonlinear. Therefore, we indirectly show the existence of the shadow Hamiltonian of $\tilde{U}_{1b}$.
Consider the following change of variables:
\begin{align}
p &= r, \\
q &= - \frac{1}{2} \log(v+1).
\end{align}
By the change of variables, the equations of motion in Eq.~(\ref{eq:eom_conserved}) are transformed to
\begin{equation}
\left\{
\begin{array}{ll}
\dot{p} & = \mathrm{e}^{-2q} -1, \\
\dot{q} & = p,
\end{array}
\right.
\end{equation}
which are Hamilton's equations of motion with the Hamiltonian
\begin{equation}
H = \frac{p^2}{2} + q +\frac{\mathrm{e}^{-2q}}{2}.
\end{equation}
We consider the first-order symplectic map
from $(p,q)$ to $(P,Q$) of this Hamiltonian,
\begin{align}
Q &= q + p h, \\
P &= p + (\mathrm{e}^{-2Q} -1) h,
\end{align}
where $P \equiv p(t+h)$ and $Q \equiv q(t+h)$ with time step $h$.
The above map is expressed in $(r,v)$-space as
\begin{align}
V &= \mathrm{e}^{- 2 r h}(v+1) - 1,\\
R &= r + V h,
\end{align}
which is identical to the map involving the propagator $\tilde{U}_{1b}$ in Eq.~(\ref{eq:vr}).
Therefore, the propagator $\tilde{U}_{1b}$ is equivalent to the first-order
symplectic integrator in $(p,q)$-space.
Since the map from $(p,q)$ to $(P,Q)$ is symplectic, 
there exists a shadow Hamiltonian $\tilde{H}(p,q)$
that satisfies $\tilde{H}(P,Q) = \tilde{H}(p,q)$
and for which the error $|H(p,q) - \tilde{H}(p,q)|$ is bounded.
This fact also guarantees the existence of a shadow Hamiltonian in $(r,v)$-space.
The relation between $(r,v)$-space and $(p,q)$-space is shown in Fig.~\ref{fig:relation}.

The time evolutions of the Hamiltonian $H$ integrated by the velocity Verlet and $~\tilde{U}_2$ schemes are shown in Fig.~\ref{fig:nh_ad}~(b). Unlike the dissipative system, the velocity Verlet algorithm conserves the Hamiltonian.
The map from $(r,v)$ to $(R,V)$ for the velocity Verlet algorithm satisfies the condition for the Jacobian in Eq.~(\ref{eq:jacobian_nh_ad}) exactly because
\begin{align}
\frac{\partial (R,V)}{\partial (r,v)} &= \frac{V+1}{v+1} \\
&= \frac{h r -1}{1 + h r +h^2 v - h^3 r(v+1)}.
\end{align}
While the above fact suggests that there exists a shadow Hamiltonian for this case, it is difficult to prove its existence.

The propagator $\tilde{U}_2$ also conserves the Hamiltonian, while the Jacobian of the transformation for $\tilde{U}_2$ is different from the exact value.
The Jacobian $J_2$ of $\tilde{U}_2$ is given by
\begin{equation}
J_2 = \exp\left( - 2rh - rvh^2 - rh^3 \right) \label{eq:J2}
\end{equation}
which is different from the exact Jacobian $J = (V+1)/(v+1)$.
Note that the inequality $J_2 \neq J$ does not mean that the integration by the propagator $U_2$ is unstable.

\section{Decomposition-order dependence} \label{sec:decomposition}

\begin{table}[tb]
\begin{tabular}{lll}
Name & Order of Operators & Order of scheme\\
\hline
Middle scheme & BAOAB & second\\
End scheme & OBAB & first \\
Beginning scheme & BABO & first \\
Side scheme & OBABO & second\\
PV-middle scheme & ABOAB & second\\
PV-end scheme & OABA & first\\
PV-beginning scheme & ABAO & first \\
PV-side scheme &OABAO & second\\
\hline
\end{tabular}
\caption{Integration schemes.
The first four correspond to velocity-Verlet and the rest to position-Verlet methods.
The label ``PV" stands for Position-Verlet.
Four of them are the first-order integration scheme and the rest are the second-order.
}
\label{tbl:shcemes}
\end{table}

\subsection{Integration Schemes}

In this section, we study the dependence of the decomposition order of the Liouville operator.
We can construct various integration schemes with Liouville operators $\mathcal{L}_A, \mathcal{L}_B,$ and
$\mathcal{L}_O$ as building blocks. 
Several previous studies indicate that the decomposition order of the Liouville operator affects the accuracy of the calculation~\cite{Itoh2013, Li2017}.
Following Lie \textit{et al.}~\cite{Li2017}, we consider the eight different decompositions which are listed in Table~\ref{tbl:shcemes}.

\subsection{Dissipative System} \label{sec:dissipative2}

\begin{figure}
\includegraphics[width=7cm]{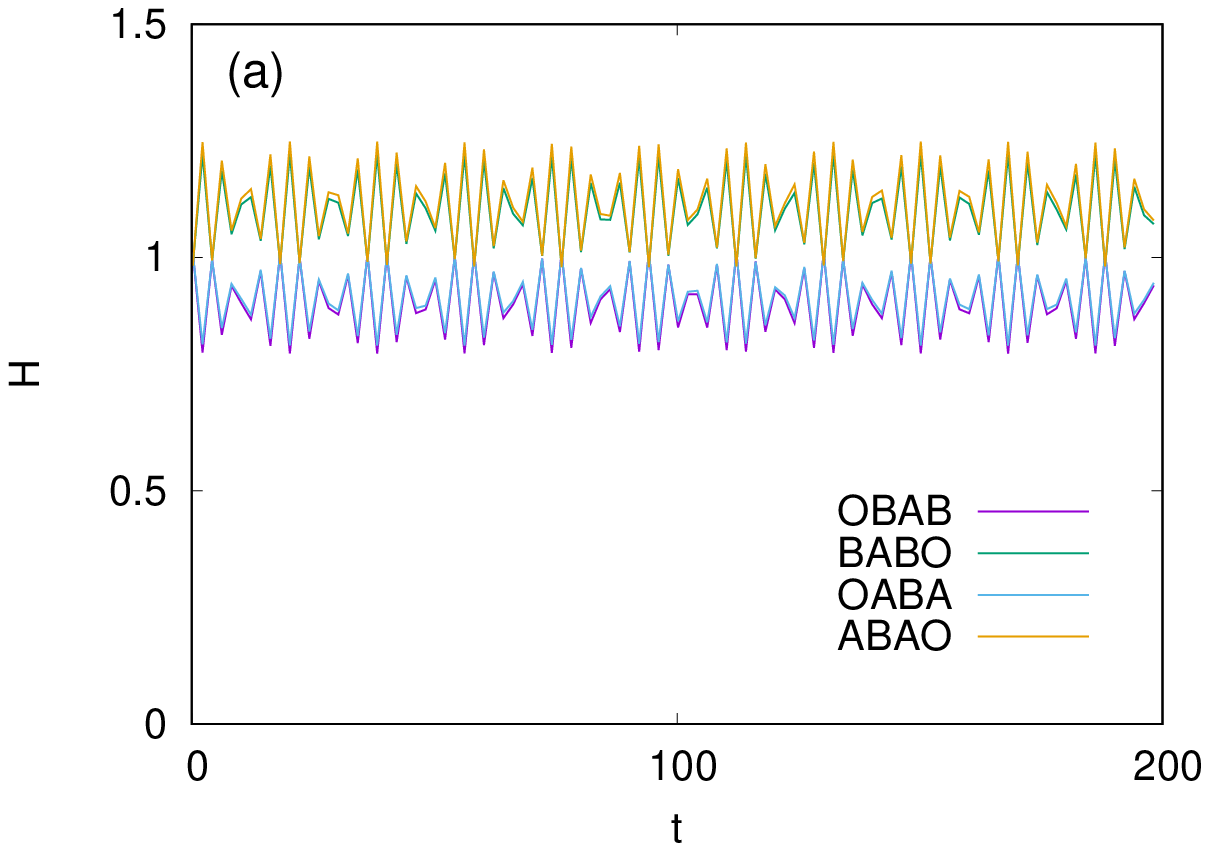}
\includegraphics[width=7cm]{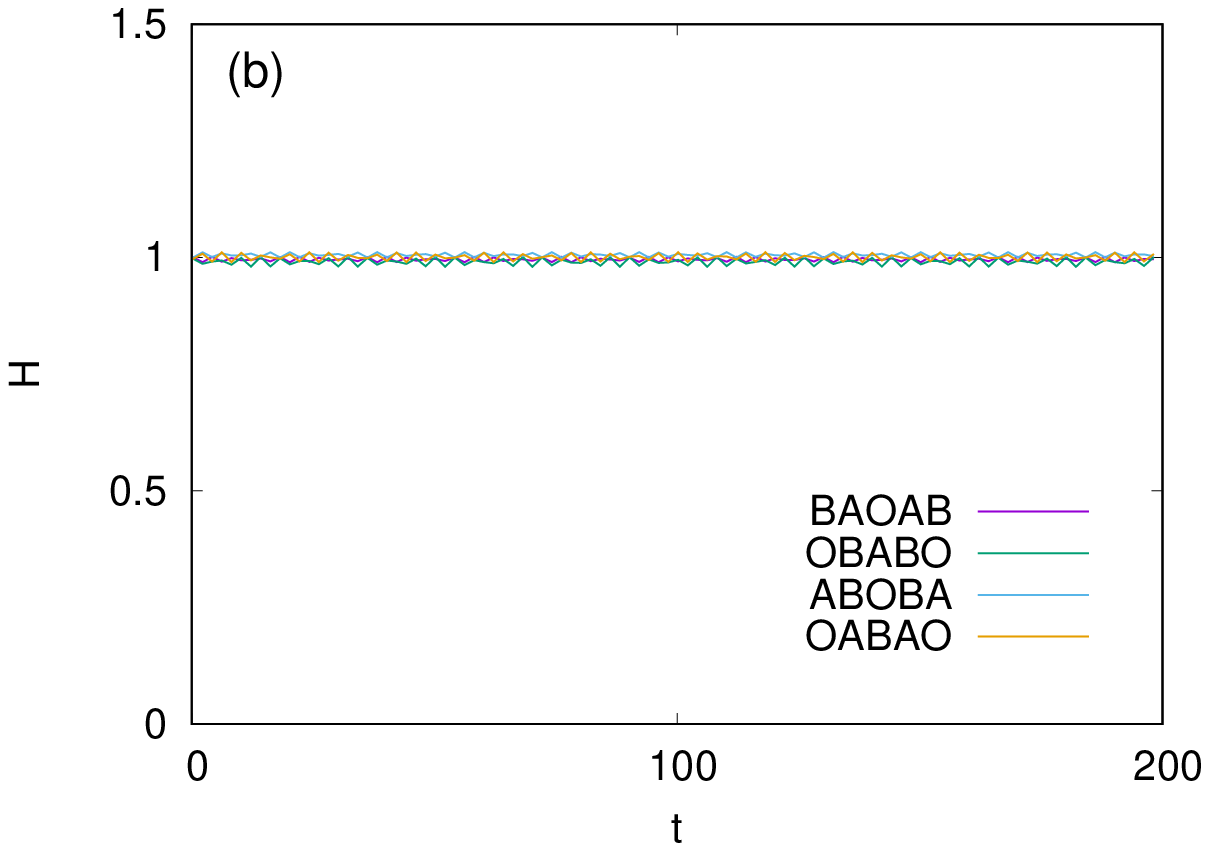}
\caption{(Color online) Decomposition-order dependence of the integration.
Time evolutions of the Hamiltonian $H$ of Eq.~(\ref{eq:conserved_value1}) are shown.
The results integrated by the first- and second-order schemes are shown in (a) and (b), respectively.
The condition of calculation is same as in Fig.~\ref{fig:dissipative}.
}\label{fig:dissipative_stability}
\end{figure}

\begin{figure}
\includegraphics[width=7cm]{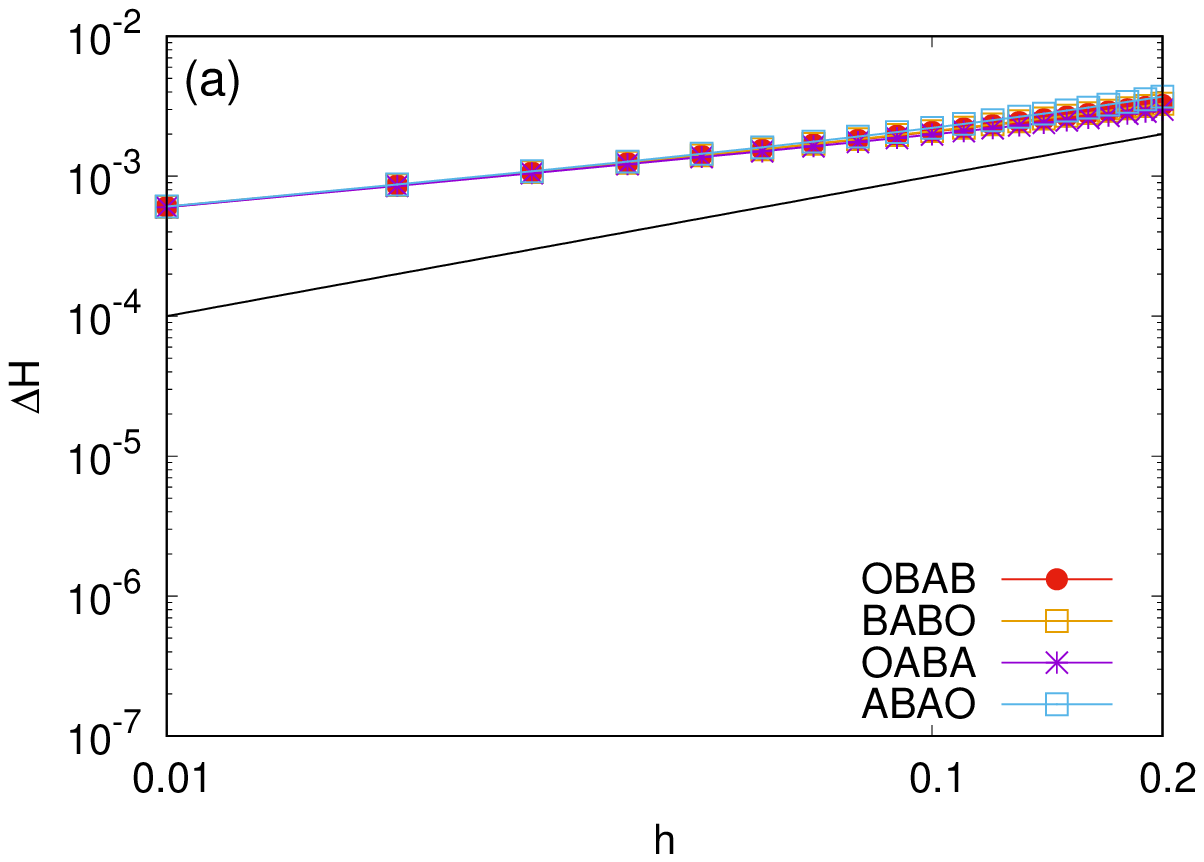}
\includegraphics[width=7cm]{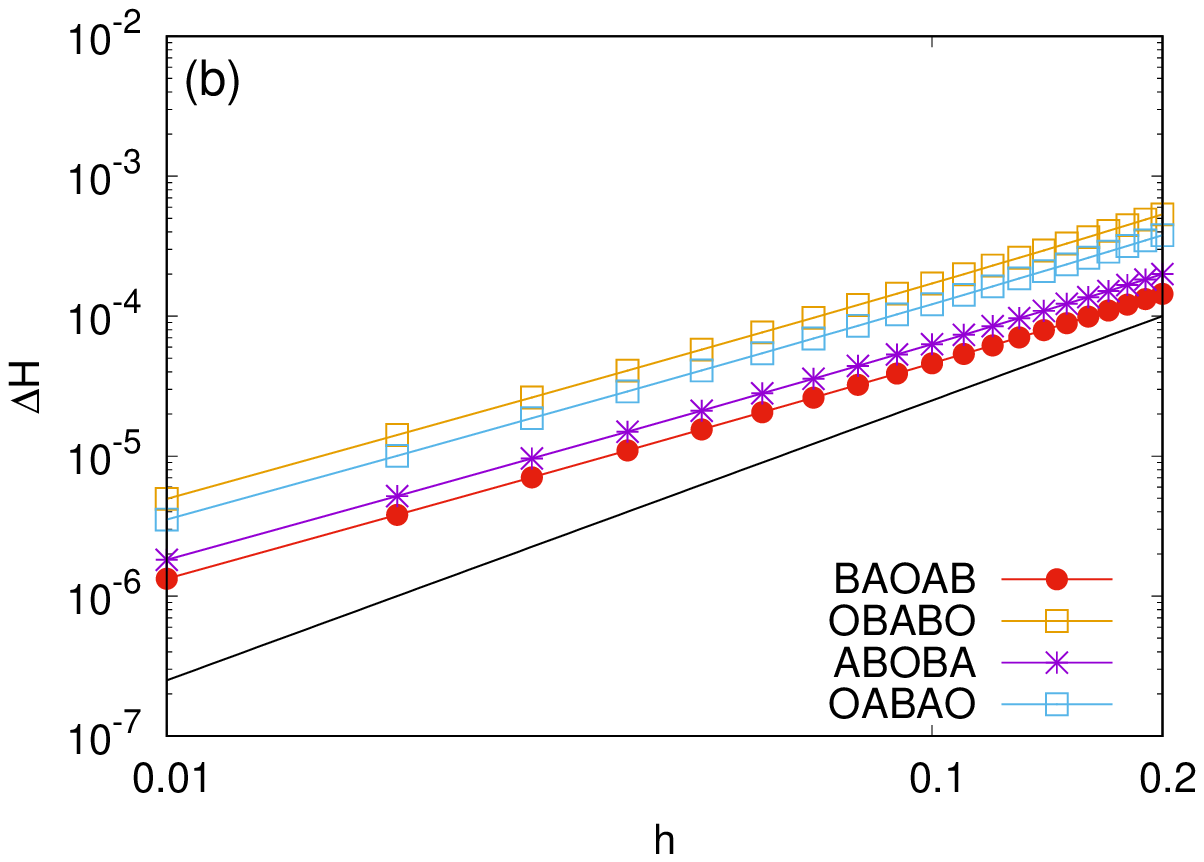}
\caption{(Color online) The accuracy of the integration schemes.
The error $\Delta H$ is shown as a function of time step $h$.
The decimal logarithms are taken for both axes. 
(a) The results integrated with the first-order schemes.
The solid line denotes $h$.
(b) The results integrated with the second-order schemes.
The solid line denotes $h^2$.
}\label{fig:dissipative_accuracy}
\end{figure}

First, we consider the dissipative system in which equations of motion is given in Eq.~(\ref{eq:dissipative}). While this system is non-Hamiltonian, we can determine the shadow Hamiltonians of the LOD schemes for this system since the system is linear.
The shadow hamiltonians of the eight schemes are as follows.
\begin{align*}
\mathrm{e}^{-h}\tilde{H}_{BAOAB} &= r^2 + \frac{2(\eh-1)rv}{h(1 + \eh)} + \left(1 - \frac{h^2}{4} \right) v^2\\
\mathrm{e}^{-h}\tilde{H}_{OBAB} &= r^2 + \frac{(1 - \eh)(h^2-2) rv}{2h \eh} + \frac{(4-h^2)v^2}{4 \eh} \\
\mathrm{e}^{-h}\tilde{H}_{BABO} &= r^2 + \frac{(1 - \eh)(h^2-2) rv}{2h } + \frac{\eh(4-h^2)v^2}{4} \\
\mathrm{e}^{-h}\tilde{H}_{OBABO} &= r^2 + \frac{\mathrm{e}^{h/2}(1 - \eh)(h^2-2) rv}{2h } + \frac{(4-h^2)v^2}{4} \\
\mathrm{e}^{-h} \tilde{H}_{ABOBA} &= \left(1 - \frac{h^2}{4} \right) r^2 - \frac{2(1 -\eh)rv}{(1 + \eh)h} + v^2\\
\mathrm{e}^{-h} \tilde{H}_{OABA} &= \frac{\eh(4-h^2)}{4} r^2 + \frac{(1 - \eh)(h^2-2)rv}{2h} +v^2 \\
\mathrm{e}^{-h} \tilde{H}_{ABAO} &=\frac{(4-h^2)r^2}{4 \eh} + \frac{(1 - \eh)(h^2-2) rv}{2h \eh} +  v^2 \\
\mathrm{e}^{-h} \tilde{H}_{OABAO} &= \frac{(4-h^2)r^2}{4} + \frac{\mathrm{e}^{h/2}(1 - \eh)(h^2-2) rv}{2h }  +v^2
\end{align*}
All of the Jacobians of the eight schemes are exactly $\mathrm{e}^{-h}$.
It is straightforward to confirm that the first or second-order schemes exhibit the first or second-order accuracy, for example, $\tilde{H}_{OBAB} - H = O(h)$ where $H = \mathrm{e}^t (r^2+rv+v^2)$.
Since the shadow Hamiltonian exists which error from the original Hamiltonian is bounded, all of these schemes are stable.
The time evolution of the Hamiltonian is shown in Fig.~\ref{fig:dissipative_stability}.
All of them are stable as expected. The fluctuations of Hamiltonian are larger with the first-order schemes compared to those with the second-order schemes.
In order to study the decomposition-order dependence on the accuracy, we observe the
error of the Hamiltonian. The error $\Delta H$ is defined as
\begin{equation}
\Delta H(h) = \sqrt{ \frac{1}{T} \sum_{k=1}^{T}\left[H(kh) - H(0)\right]^2},
\end{equation}
where $T$ is a number of total steps.
We performed integrations for 1000 steps with various time steps to evaluate the errors.
The errors are shown in Fig.~\ref{fig:dissipative_accuracy}.
While the second-order schemes exhibit $O(h^2)$ behavior error clearly, 
the first-order schemes exhibit deviations from $O(h)$ behavior.
The most accurate scheme is BAOAB type as reported previously.~\cite{Leimkuhler2013}.

\subsection{Conserved System}

\begin{figure}
\includegraphics[width=7cm]{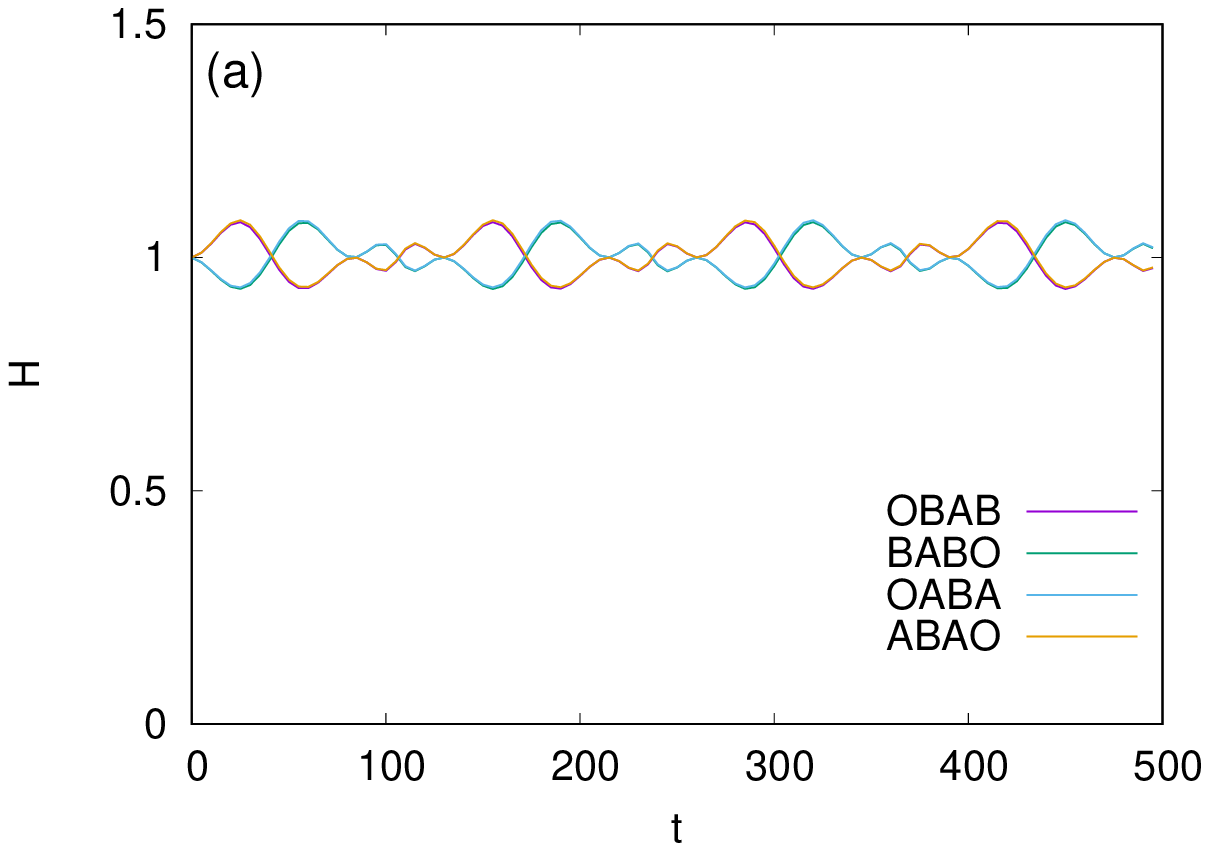}
\includegraphics[width=7cm]{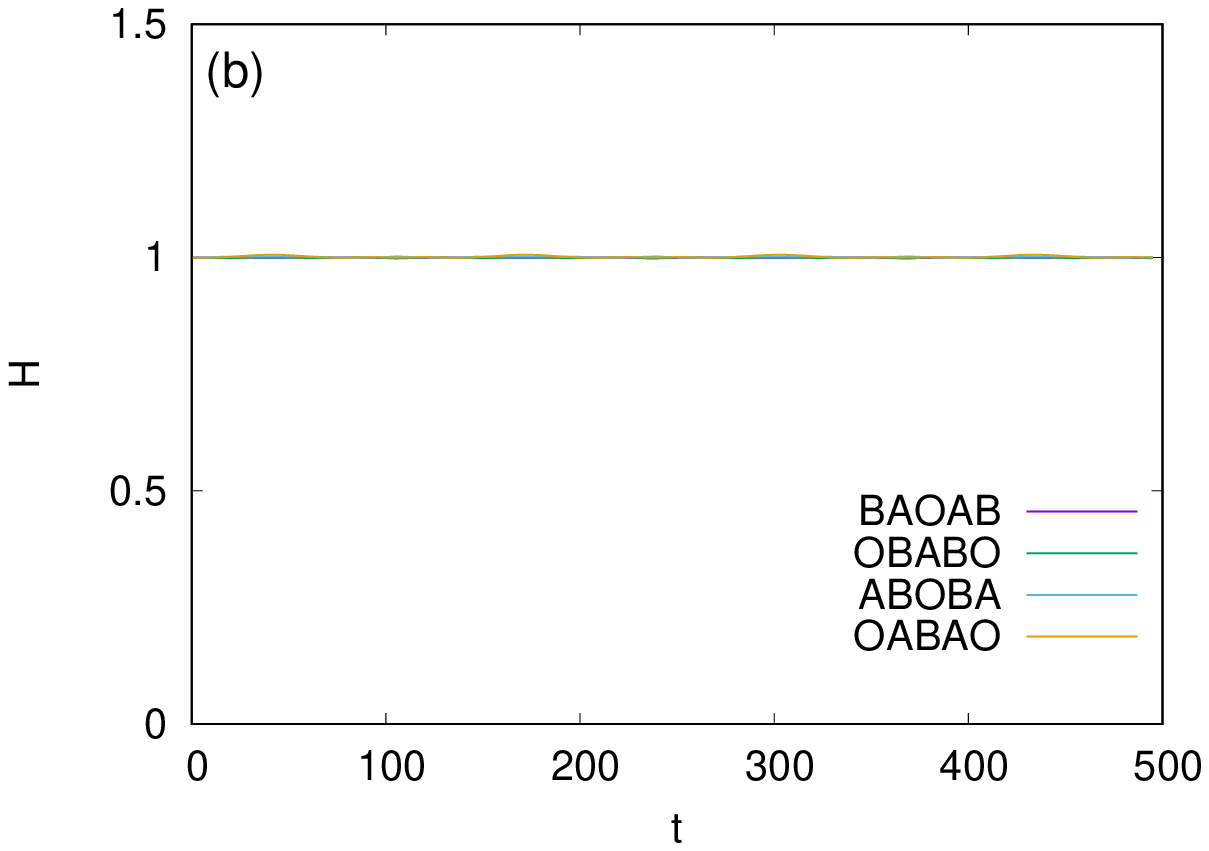}
\caption{(Color online) Decomposition-order dependence of the integration.
Time evolutions of the Hamiltonian $H$ of Eq.~(\ref{eq:H_conserved}) are shown.
The results integrated by the first- and second-order schemes are shown in (a) and (b), respectively.
The condition of calculation is same as in Fig.~\ref{fig:nh_ad}.
}\label{fig:conserved_stability}
\end{figure}

For the conserved system which equation of motion is given in Eq.~(\ref{eq:eom_conserved}),
we cannot determine the shadow Hamiltonian in the closed form since the system is non-linear.
The time evolutions of the Hamiltonian integrated with the eight schemes are shown in Fig.~\ref{fig:conserved_stability}.
All of them are stable, while the first-order propagator studied in Sec.~\ref{sec:conserved_system} are unstable.

\begin{figure}
\includegraphics[width=7cm]{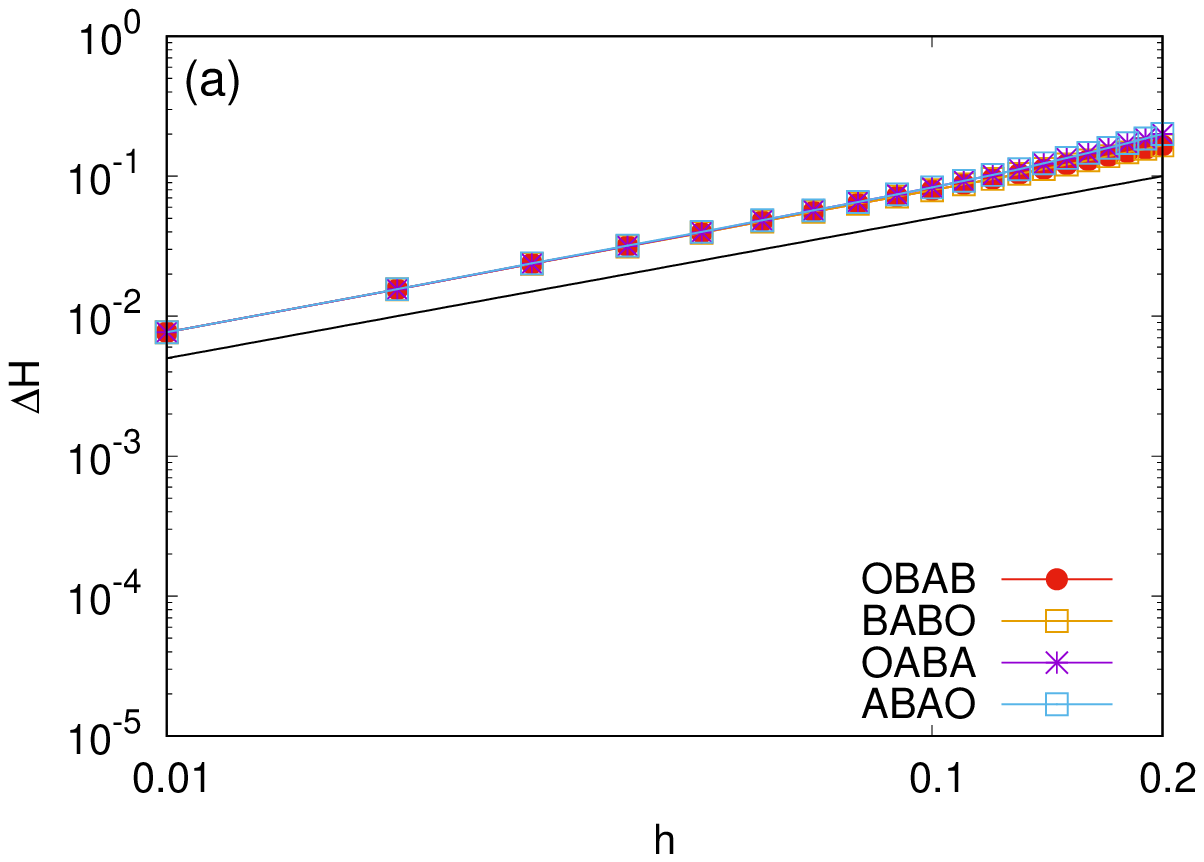}
\includegraphics[width=7cm]{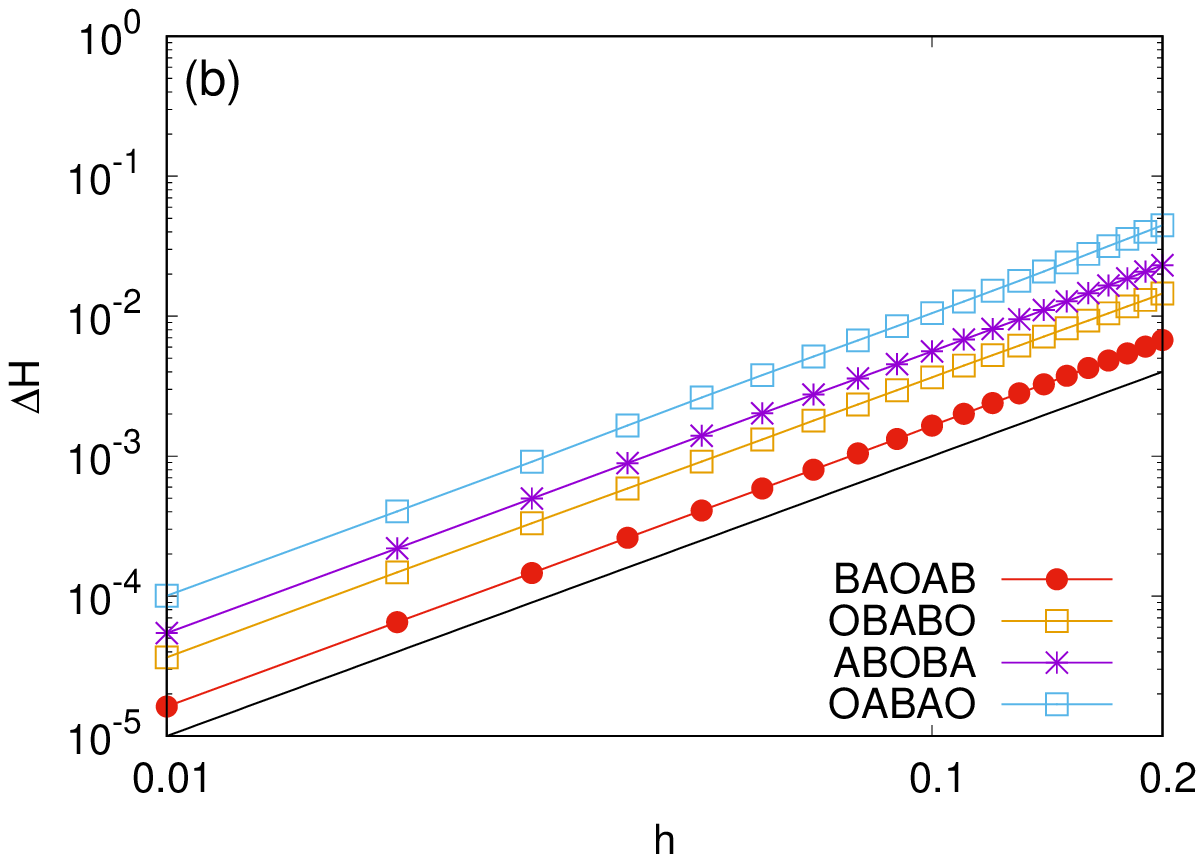}
\caption{(Color online) Accuracy of the integration schemes.
The error $\Delta H$ is shown as a function of time step $h$.
The decimal logarithms are taken for both axes. 
(a) The results integrated with the first-order schemes.
The solid line denotes $h$.
(b) The results integrated with the second-order schemes.
The solid line denotes $h^2$.
}\label{fig:conserved_accuracy}
\end{figure}

The error of the conserved value is shown in Fig.~\ref{fig:conserved_accuracy}.
One can confirm that the first- and second-order behaviors for the first- and second-order schemes.
Again, the BAOAB scheme exhibits the best accuracy.

\section{Summary and Discussion}\label{sec:summary}

In the present article, we have investigated two non-Hamiltonian systems for which the Jacobian can be determined exactly. One is a dissipative system with a time-dependent conserved value (the first system),
and the other is a non-Hamiltonian system with a time-independent conserved value (the second system).
For the first system, the velocity Verlet algorithm failed to conserve the Hamiltonian,
while the first- and the second-order LOD algorithms conserved it.
We have given proofs why and how the velocity Verlet algorithm fails and the LOD algorithms succeed in conserving the Hamiltonian.
Meanwhile, the stability of the integrators applied to the second system is different from that for the first system. The velocity Verlet and second-order LOD algorithms conserve the Hamiltonian while
the first-order LOD algorithm does not.
We have constructed an improved version of the first-order LOD scheme by adopting a different factorization of the propagator, and the improved algorithm conserves the Hamiltonian.
Since both first-order algorithms are time-irreversible, the failure of the first-order LOD algorithm
does not originate from the time-irreversibility.
We have proved that the value of the Jacobians of the failed schemes is always less than the exact values.
Then the values of the phase space decrease monotonically, and consequently, the Hamiltonians are not conserved.
It is found that the second-order LOD is stable both for the dissipative and conserved systems,
but the reason why it is stable is not given.
It should be addressed in the future.

The decomposition-order dependence is also studied.
We have considered eight integration schemes. As reported previously,
BAOAB type exhibits the best accuracy.
The first-order propagator $\tilde{U}_{1a}$ in Eq.~(\ref{eq:UABO}) studied in Sec.~\ref{sec:conserved_system} corresponds to ABO scheme.
While the ABO scheme is unstable for this system, the BABO and other first-order schemes studied here are found to be stable. This result suggests that the integration scheme is stable when the propagator of the unitary part is of the second-order. However, we do not have any proof of this. This is also one of the further issues.

\section*{Acknowledgements}
We would like to thank N. Kawashima and H. Noguchi
for helpful discussions.
This work was supported by JSPS KAKENHI Grant Number 15K05201 and
by MEXT as ``Exploratory Challenge on Post-K computer'' (Challenge of Basic Science-Exploring Extremes through Multi-Physics and Multi-Scale Simulations).

\section*{Appendix A}

A symplectic integrator conserves a value that is very close to the original Hamiltonian.
The conserved value is often called a shadow Hamiltonian.
While it is difficult to determine the exact form of a shadow Hamiltonian, we can obtain it when the propagator is linear.
Consider the following harmonic oscillator:
\begin{equation}
\left\{
\begin{array}{ll}
\dot{p} &= -q,\\
\dot{q} &= p.
\end{array}
\right.
\end{equation}
The Hamiltonian is $H = p^2/2 + q^2/2$.
We want to approximate the propagator of this system $U(h)$, which proceeds the time by $h$.
We consider the first- and second-order approximations $\tilde{U}_1(h)$ and $\tilde{U}_2(h)$, respectively.
They are given by the factorization of the propagators as
\begin{align}
\tilde{U}_1(h) &= \exp{\left(-h q \frac{\partial}{\partial p}\right)} \exp{\left(h p \frac{\partial}{\partial q}\right)},\\
\tilde{U}_2(h) &= \exp{\left( \frac{h}{2} p \frac{\partial}{\partial q}\right)}\exp{\left(-h q \frac{\partial}{\partial p}\right)} \exp{\left( \frac{h}{2} p \frac{\partial}{\partial q}\right)}.
\end{align}
Since the system is linear, the propagators can be expressed in a matrix form as
\begin{align}
\tilde{U}_1 &= 
\begin{pmatrix}
1-h^2 & -h \\
h & 1
\end{pmatrix}, \\
\tilde{U}_2 &= 
\begin{pmatrix}
1-h^2/2 & -h \\
h - h^3/4 & 1 - h^2/2
\end{pmatrix}. \\
\end{align}
The original Hamiltonian of this system can be expressed by the quadratic form
\begin{equation}
2 H(p,q) = p^2 + q^2 = 
(p, q) I
\begin{pmatrix}
p \\
q
\end{pmatrix},
\end{equation}
where $I$ is the identity matrix. 
Therefore, we assume that the shadow Hamiltonian $\tilde{H}_1$ also has the quadratic form
\begin{equation}
\tilde{H}_1(p,q) = (p, q) X_1
\begin{pmatrix}
p \\
q
\end{pmatrix} \label{eq:shadow1}
\end{equation}
with $2 \times 2$ matrix $X_1$.
Since the shadow Hamiltonian is exactly conserved by the propagator $\tilde{U}_1$, the identity
\begin{equation}
\tilde{H}_1(P,Q) = \tilde{H}_1(p,q) \label{eq:shadow2}
\end{equation}
is satisfied, where
\begin{equation}
\begin{pmatrix}
P \\
Q
\end{pmatrix} = 
\tilde{U}_1
\begin{pmatrix}
p \\
q
\end{pmatrix}. \label{eq:shadow3}
\end{equation}
From Eqs.~(\ref{eq:shadow1}), (\ref{eq:shadow2}), and (\ref{eq:shadow3}), we 
have the following condition for $X_1$:
\begin{equation}
\tilde{U}_1^{T} X_1 \tilde{U}_1 = X_1.
\end{equation}
From the fact that $\lim_{h\rightarrow 0} X_1 = I$, we have
\begin{equation}
X_1 = 
\begin{pmatrix}
1 & h \\
0 & 1
\end{pmatrix}.
\end{equation}
Thus, the shadow Hamiltonian $\tilde{H}_1$ conserved by $\tilde{U}_1$ is determined to be
\begin{equation}
H_1(p,q) = \frac{1}{2}(p^2 - hpq + q^2).
\end{equation}
One can confirm that the identity $H_1(P,Q) = H_1(p,q)$ exactly holds.
Similarly, we find that the shadow Hamiltonian $\tilde{H}_2$ conserved by $\tilde{U}_2$ is $2 \tilde{H}_2(p,q) = (1-h^2/4) p^2 + q^2$.
Parallel arguments lead to the exactly conserved values in the dissipative system described in Sec.~\ref{sec:dissipative_system} and ~\ref{sec:dissipative2}.

\section*{Appendix B}

Consider the following equations of motion,
\begin{align}
\dot{p} &= -q - p \zeta, \label{eq:dotp} \\
\dot{q} &= p, \\
\dot{\zeta} &= \frac{1}{Q}(p^2 - 1).\label{eq:dotzeta}
\end{align}
The above equations of motions represent a harmonic oscillator with
the Nos\'e--Hoover thermostat. The Boltzmann constant, the mass of the oscillator, and the target temperature are set to unity. The fictitious mass of the thermostat is denoted by $Q$.
The second order differential of $\zeta$ is
\begin{equation}
\ddot{\zeta} = \frac{2 p \dot{p}}{Q} \label{eq:ddotzeta}.
\end{equation}
Substituting Eqs.~(\ref{eq:dotp}) and (\ref{eq:dotzeta}) into Eq.~(\ref{eq:ddotzeta}), we have
\begin{equation}
\ddot{\zeta} = 2 \zeta \dot{\zeta} + \frac{2 \zeta}{Q} - \frac{2 pq}{Q}.
\end{equation}
For sufficiently large $Q$, $\zeta$ varies much more slowly than $p$ and $q$.
Since the value of $pq$ oscillates around zero, we can replace $pq$ with $0$ (adiabatic approximation). Then we have
\begin{equation}
\ddot{\zeta} = 2 \zeta \dot{\zeta} + \frac{2 \zeta}{Q}.
\end{equation}
Applying the transformation $r = \sqrt{Q}^{-1} \zeta(t/\sqrt{Q})$ and $\dot{r} = v$,
we have
\begin{equation}
\left\{
\begin{array}{ll}
\dot{v} &= - 2 r - 2 r v, \\
\dot{r} &= v,
\end{array}
\right.
\end{equation}
which are the equations of motion considered in Sec.~\ref{sec:conserved_system}.

\bibliography{velocityverlet}

\end{document}